\documentclass[final,5p,times,twocolumn]{elsarticle}
\usepackage{amsthm,amsmath,amssymb}
\usepackage{xspace}

\title{Modeling wind farm noise emission and propagation: effects of flow and layout}

\author[1]{Jules Colas\corref{cor1}}
\author[1]{Ariane Emmanuelli}
\author[1]{Didier Dragna}
\author[2]{Richard J. A. M. Stevens}

\affiliation[1]{Ecole Centrale de Lyon, CNRS, Universite Claude Bernard Lyon 1, INSA Lyon, LMFA, UMR5509, 69130,
Ecully, France}
\affiliation[2]{Physics of Fluids Group, Max Planck Center Twente for Complex Fluid Dynamics, J. M. Burgers Center for Fluid Dynamics, University of Twente, P.O. Box 217, 7500 AE Enschede, The Netherlands}
\cortext[cor1]{Corresponding author: jules.colas@ec-lyon.fr}

\journal{Renewable Energy}

\newcommand{\OASPL}{$\overline{\rm OASPL}$\xspace}
\newcommand{\ms}{m.s$^{-1}$\xspace}
\newcommand{\dl}{$\Delta L$\xspace}

\renewcommand{\aa}{(a)\xspace}
\newcommand{\bb}{(b)\xspace}
\newcommand{\cc}{(c)\xspace}
\newcommand{\dd}{(d)\xspace}
\newcommand{\ee}{(e)\xspace}
\newcommand{\ff}{(f)\xspace}
\renewcommand{\gg}{(g)\xspace}
\newcommand{\hh}{(h)\xspace}
\newcommand\fracc[2]{\frac{\displaystyle #1}{\displaystyle #2}}

\journal{Renewable Energy}
\begin{document}
\begin{abstract}
This study demonstrates how wind farm flow physics influence noise generation and downstream propagation through numerical simulations. 
The flow field is modeled using large-eddy simulations (LES), and the time-averaged output serves as input to acoustic models that predict wind turbine noise.
In the first turbine row, turbulence-induced noise (TIN) and trailing edge noise (TEN) contribute equally, with TIN dominating at low frequencies and TEN at higher frequencies.
Downstream, TEN decreases due to lower wind speeds, while TIN mostly persists due to increased turbulence dissipation. 
These effects are more pronounced in aligned wind farms, where stronger wake interactions occur, than in staggered layouts. 
However, staggered farms produce more noise overall because turbines operate at higher wind speeds.
Additionally, wind farm flow significantly affects sound propagation downwind.
The wake superposition  modifies sound focusing leading to different amplification area than for an isolated turbine. 
For a staggered layout it particularly shows enhanced sound focusing downwind due to the position of the turbine wakes. 
This leads to higher sound levels and higher amplitude modulation downwind for the wind farm compared to an aligned layout.
These phenomena are not captured by models based on isolated turbines.
These findings underscore the importance of integrating flow and acoustic models to more accurately assess the environmental impact of wind farms.
\end{abstract}


\maketitle

%%%%%%%%%%%%%%%%%%%%%%%%%%%%%%%%%%%%%%%%%%%%%%%%%%%%%%%%%%%%%%%%%%%%%%%%%%%%%%%
%%%%%%%%%%%%%%%%%%%%%%%%%%%%%%%%%%%%%%%%%%%%%%%%%%%%%%%%%%%%%%%%%%%%%%%%%%%%%%%
\section{Introduction}
The expansion of renewable energy has led to an increase in both onshore and offshore wind farms, which consist of multiple wind turbines operating in close proximity.
The number of onshore wind farm is expected to keep increasing to reach Europe's energy production goals \citep{costanzoWindEnergyEurope2024}.
Hence, assessing wind turbine noise through simulation is crucial to mitigate its effects on residential areas. 
For accurate assessment, it is essential to consider both realistic sound power levels (SWL) and propagation effects. 
Previous studies have examined the effect of flow on sound propagation for an isolated turbine, and described a strong influence of atmospheric conditions on sound propagation, as well as an effect of the wind turbine wake on downwind propagation, leading to increased sound pressure level (SPL) in this direction \citep{barlasEffectsWindTurbine2017, heimann3DsimulationSoundPropagation2018, kayserEnvironmentalParametersSensitivity2020,bommidalaThreedimensionalEffectsWake2025}.

However, flow conditions inside and around a wind farm can differ significantly from those around an isolated wind turbine, affecting both noise generation and propagation.
More over, large wind farm can lead to the formation of an internal boundary layer, which modifies the atmospheric flow downwind \citep{porte-agelWindTurbineWindFarmFlows2020,stevensFlowStructureTurbulence2017}.
% In this chapter, the case of multiple turbine interactions is addressed and the specific challenges that come with it.
Some studies have already contributed to the understanding of wind farm noise propagation.
\citet{shenAdvancedFlowNoise2019} performed a detailed investigation of noise propagation in wind farms using parabolic equation (PE) simulation, accounting for realistic atmospheric flow and terrain features.
Their study demonstrated that variations in the mean flow lead to substantial changes in sound pressure levels upstream and downstream, relative to those predicted using a logarithmic profile.
\citet{sunDevelopmentEfficientNumerical2018} also used PE simulation for wind farm noise prediction and studied the effect of wake superposition in the downwind direction.
They concluded that there is a combined effect of atmospheric conditions and wake effect on the downwind and upwind noise levels of a wind farm.
Their study was limited to downstream and upstream propagation directions and considered a simple source model.
Finally, \citet{nyborgOptimizationWindFarm2023} used complete wind farm simulations accounting again for the wind turbines and their wake to investigate the optimization of curtailment plans for noise reduction.
They show that accounting for realistic noise propagation (with shear and wake effects) could lead to more efficient curtailment plans. 

Despite this previous work, there is a lack of comprehensive study on wind farm noise propagation. 
Specifically, there has been a lack of comparison between different wind farm layouts and extensive analysis of the potential effects arising from propagation inside a wind farm has not yet been conducted.
This work aims to address these gaps by conducting a comparative study of wind farm noise propagation for different wind farm layouts.
This is crucial to provide recommendations for the optimal layout of wind farms to minimize noise impacts.

The paper is organized as follow.
Sec.~\ref{s: methods} details the numerical methods employed to compute the noise from a wind farm. 
The different wind farm layouts are then introduced in Sec.~\ref{s: cases}.
Initially, an isolated turbine case with a stable atmospheric boundary layer (ABL) is considered.
Subsequently, aligned and staggered wind farm layouts are presented.
The flow fields obtained for these three cases are shown in Sec.~\ref{s: flow}.
Secs.~\ref{s: source} and \ref{s: deltaL} study the influence of the layout on the source levels and on the propagation, respectively. 
In Sec.~\ref{s: oaspl}, the overall sound pressure level (OASPL) and amplitude modulation (AM) fields are compared between the different cases.
Finally, concluding remarks are given in Sec.~\ref{s: conclusion}.

%%%%%%%%%%%%%%%%%%%%%%%%%%%%%%%%%%%%%%%%%%%%%%%%%%%%%%%%%%%%%%%%%%%%%%%%%%%%%%%
%%%%%%%%%%%%%%%%%%%%%%%%%%%%%%%%%%%%%%%%%%%%%%%%%%%%%%%%%%%%%%%%%%%%%%%%%%%%%%%
\section{Methodology}
\label{s: methods}

The methodology used to compute wind turbine noise involves three main steps.
This approach is consistent with the methodology outlined in \citet{colasWindTurbineSound2023} and is summarized briefly below.

\subsection{Flow}
First, large-eddy simulations (LES) \citep{gaddeLargeEddySimulationsStratified2021,gaddeEffectCoriolisForce2019,stierenModelingDynamicWind2021, stevensConcurrentPrecursorInflow2014} are performed to obtain the average flow field.
The wind turbines are modeled with an actuator disk method \citep{stevensComparisonWindFarm2018}. 
Although the simulations capture unsteady dynamics, this study focuses only on the mean velocity fields $\mathbf{V_0}=(u_0, v_0, w_0)$ and turbulence dissipation rate $\epsilon_0$, excluding turbulence scattering and unsteady effect even though it is known to impact wind turbine noise propagation \citep{barlasEffectsWindTurbine2017}.

\subsection{Source model}
The noise generated by each wind turbine is computed using the approach described in \citet{tianWindTurbineNoise2016}. 
Initially, the free-field sound pressure level SPL$_{ \rm ff}$ is calculated based on Amiet strip theory \citep{amietNoiseDueTurbulent1976}.
Each turbine blade is divided into several segments, treated as uncorrelated point sources. 
The trailing edge noise (TEN)  and turbulent inflow noise (TIN)  are computed for each segment at various angular positions, accounting for the blade geometry and the input mean flow \citep{tianWindTurbineNoise2016,
mascarenhasSynthesisWindTurbine2022}. 
The TIN depends on the local wind speed and turbulence dissipation rate encountered by each blade segment.
The local wind speed at each blade segment is determined from the blade rotational speed $\Omega$ and the incoming wind speed $u_0$.
The turbulence dissipation rate $\epsilon_0$ is computed directly from the LES results at each blade segment position. 
The TEN depends on the local wind speed and airfoil geometry.
The wall pressure fluctuation at the trailing edge is computed using the model developed by \citet{leePredictionAirfoilTrailingEdge2019} with boundary layer quantities computed using XFoil. 
The TIN and TEN models are applicable to the far field, defined as distances greater than the blade chord length and the acoustic wavelength. 

In this work the data provided for the NREL 5~MW wind turbine \citep{jonkmanDefinition5MWReference2009} is used to control $\Omega$ as a function of the wind speed sampled at the hub height ($u_{\rm hub}$) of each turbine.
The rotor speed increases from the cut-in wind speed to the rated where it is kept constant up to the cut-out wind speed (see Table~\ref{t: control}).
\begin{table}[h!]
    \caption{Wind turbine control parameters \citep{jonkmanDefinition5MWReference2009}}
    \label{t: control}
\centering
 \begin{tabular}{c c c c} 
 \hline
 \hline
  & cut-in & rated & cut-out \\
 \hline
     $u_{\rm hub}$ (\ms) & 3 & 11.4 & 25 \\ 
 $\Omega$ (rpm) & 6.9 & 12.1 & 0\\
 \hline
 \hline
 \end{tabular}
\end{table}
\subsection{Propagation model}
%% Propagation model 
%%%%%%%%%%%%%%%%%%%%%%%%%%%%%%%%%%%%%%%%%%%%%%%%%%%%%%%%%%%%%%%%%%%%%%%%%%%%%%%
The sound pressure level relative to the free-field $\Delta L$ is computed using a wide angle parabolic equation model (WAPE) which formulation is derived in  Sec.~V.A. of \citet{ostashevWaveExtrawideangleParabolic2020} for an arbitrary moving medium using a high frequency approximation.
It accounts for ground effects and atmospheric refraction on sound propagation. 
The WAPE is solved using a second-order finite-difference scheme in the $z$-direction and a Crank–Nicolson algorithm to advance the solution  in the $x$-direction.
Numerically, this involves inverting a tridiagonal matrix, which is done efficiently using the Thomas algorithm. The starting field is the second-order starter presented in \citet{salomonsComputationalAtmosphericAcoustics2001}, which represents a monopole source.

To optimize computational costs, $\Delta L$ is not calculated for each segment position. 
Instead, fictive source heights are assumed along a vertical line in the rotor plane passing through the turbine hub as described in \citet{cotteExtendedSourceModels2019}.
The value of $\Delta L$ is interpolated linearly between the two closest fictive sources.
For each turbine in the wind farm, simulations are performed for multiple propagation angles $\tau$. 
The resulting 2D fields are interpolated onto a common 3D Cartesian grid.

\subsection{Sound pressure level and amplitude modulation}
The SPL at a receiver for one segment is given by
\begin{equation}\label{e:salomons}
	\begin{aligned}
		\operatorname{SPL}^i(\mathbf{x}, f, \beta)= & \operatorname{SPL}_{\mathrm{ff}}^i(\mathbf{x}, f, \beta) +\Delta L^i(\mathbf{x}, f, \beta)-\alpha(f) R,
	\end{aligned}
\end{equation}
where $i$ denotes the segment index, $\mathbf{x}$ the receiver coordinates, $f$ the frequency, $\beta$ the blade angle ($\beta=0$ corresponds to the blade pointing upwards), $\alpha$ the atmospheric absorption coefficient, and $R$ the distance from source to receiver.

% Time domain 
%%%%%%%%%%%%%%%%%%%%%%%%%%%%%%%%%%%%%%%%%%%%%%%%%%%%%%%%%%%%%%%%%%%%%%%%%%%%%%%
The noise emitted by each segment is considered fully uncorrelated such that the SPL resulting from each blade segment can be summed logarithmically to obtain the sound from a full wind turbine.
Similarly, in the case of a wind farm, the contributions of each turbine can be added logaritmically. 
When considering multiple turbines at different location, the propagation time must also be accounted for.
Following the methodology from \citet{mascarenhasSynthesisWindTurbine2022} the SPL as a function of the rotor angle is converted to an SPL as a function of time.
The rotor is discretized with an angular step $\Delta \beta$ and the time domain is discretized with a time step $\Delta t$.
Time signals of length $T_{\rm tot} = 2\pi / \Omega$ are computed at each receiver position such that
\begin{equation}
	\mathrm{SPL}(\mathbf{x}, t_k, f) = 10\log_{10} \left( \sum_{i=1}^{N_s} \sum_{j=1}^{N_\beta} 10^{{\rm SPL}^i(\mathbf{x}, \beta_j, f)} \delta_{ijk}\right),
\end{equation}
\begin{equation}
	\delta_{ijk} = \begin{cases}
        1, & \text{if} \  T_i^{j+1} < t_k < T_i^j \\
		0, & \text{otherwise}
	\end{cases}
	, \qquad
	T_i^j = \fracc{R_i^j(\mathbf{x})}{c_{\rm ref}} + \fracc{\beta_j}{\Omega},
\end{equation}
where $\Omega$ is the rotational speed in rad.s$^{-1}$, $T_i^j$ is the time at which the sound emitted by a segment $i$  at a position $\beta_j=j\Delta \beta$ reaches a receiver at $\mathbf{x}$, $t_k=k\Delta t$ is the discrete time, $R_i^j(\mathbf{x})$ is the distance between segment $i$ at angle $\beta_j$ and receiver $\mathbf{x}$, and $c_{\rm ref}$ is the sound speed at the ground.
Note that this is an approximation as the sound speed is not consistent in the entire domain.
A more accurate approximation, which was not implemented, would involve using the average sound speed along the path between each source-receiver pair.
Note that if two turbines rotate at different speeds $\Omega$ the total time computed is different and hence a common multiple of the two periods must be used to recover a periodic signal.

Finally, the overall sound pressure level (OASPL) is obtained by integrating the SPL across the frequency band \citep{colasWindTurbineSound2023}.
The averaged OASPL ($\overline{\rm OASPL}$) and AM are computed over a complete rotor rotation as:
\begin{equation}
	\begin{aligned}
         & \overline{\rm OASPL}(\mathbf{x}) = 10 \log_{10} \left( \fracc{2\pi}{\Omega} \sum_k^{N_t} \Delta t 10^{{\rm OASPL}(\mathbf{x},t_k)/10}  \right),                     \\
         & {\rm AM}(\mathbf{x}) =  \max_{t}({\rm OASPL}(\mathbf{x},t)) - \min_t({\rm OASPL}(\mathbf{x},t)),
	\end{aligned}
\end{equation}
where $N_t$ is the number of discrete time step for one rotor rotation.

%%%%%%%%%%%%%%%%%%%%%%%%%%%%%%%%%%%%%%%%%%%%%%%%%%%%%%%%%%%%%%%%%%%%%%%%%%%%%%%
%%%%%%%%%%%%%%%%%%%%%%%%%%%%%%%%%%%%%%%%%%%%%%%%%%%%%%%%%%%%%%%%%%%%%%%%%%%%%%%
\section{Cases studied}
\label{s: cases}
\begin{figure}[htb]
    \centering
    \footnotesize
    \includegraphics{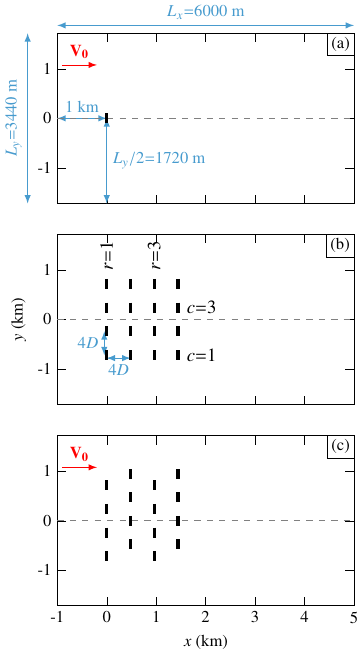}
    \vspace{-0.3cm}
    \caption{Sketch of the LES domain for the different layouts. Wind turbines are indicated using thick lines. Note that the acoustic computational domain only extends up to $x=5000~$m.}
    \label{f6: config}
\end{figure}
Different scenarios are examined with a computational domain of $L_x=6~$km, $L_y=3.44~$km, and $L_z=1~$km in the streamwise, spanwise, and vertical directions, see Fig.~\ref{f6: config}.
% turbine geometry and position 
To provide a reference, simulations are performed with a single wind turbine and compared to aligned and staggered wind farm configurations comprising 4 $\times$ 4 turbines.
The turbines have a diameter of $D=120~$m and a hub height of $h_{\rm hub}=90~$m. 
In the aligned wind farm case, the inter-turbine spacing in the spanwise and streamwise direction is $4D~=~480~$m.
The thrust coefficient and induction factor for the actuator disk model are $C_T=3/4$ and $a=1/4$, respectively.
Note that the columns of the wind farm are aligned with the flow direction while the rows are considered perpendicular to the flow.
In the following, each row and column are referenced with index $r$ and $c$, respectively.
Hence, turbines are designated as $t_r^c$, such that $t_1^1$ corresponds to the turbine at $(x,y)=(0,-720)~$m.

The LES domain is discretized using a grid of $400\times 220 \times 96$ points, yielding a resolution of $\Delta x~=~15~$m, $\Delta y~=~15.6~$m and $\Delta z~=~10.4~$m.
The concurrent precursor method \citep{stevensConcurrentPrecursorInflow2014} is employed with a fringe region length of $600~$m in the streamwise direction.
The surface roughness is set to $z_0=0.1~$m, a typical value used in acoustics for onshore conditions \citep{salomonsComputationalAtmosphericAcoustics2001}.
Stable boundary layer cases are simulated with a constant cooling rate equal to $C_L=-0.2~$K.h$^{-1}$ applied at the ground.
Simulations are run until a quasi-equilibrium state is reached, after which the flow fields are averaged over two hours to provide input for the source and propagation models. 

% acoustic simulation numerical parameters 
Acoustic simulations are performed for each case, at frequencies ranging from 50~Hz to 1080~Hz with a grid step equal to a tenth of the wavelength, consistent with \citet{colasImpactTwodimensionalSteep2024}.
The wind turbine geometry is similar to the one developed by \citet{cotteExtendedSourceModels2019}.
The rotor is discretized with an angular step of 10°, and each blade is divided into 8 segments.
The propagation simulation is carried out using the N$\times$ 2D-PE propagation method.
An angular step of 2° is used to discretize the entire domain except in the downwind direction ($-16$°$<\tau<16$°) where a 1° angular step is employed. 
The results are then interpolated on a Cartesian grid with $\Delta x=\Delta y=10~$m.
The propagation simulations are performed for 7 equivalent source heights to discretize the rotor plane \citep{colasImpactTwodimensionalSteep2024,colasWindTurbineSound2023}.

\section{Flow fields}
\label{s: flow}
\begin{figure*}[htbp]
    \footnotesize
    \centering
    \includegraphics{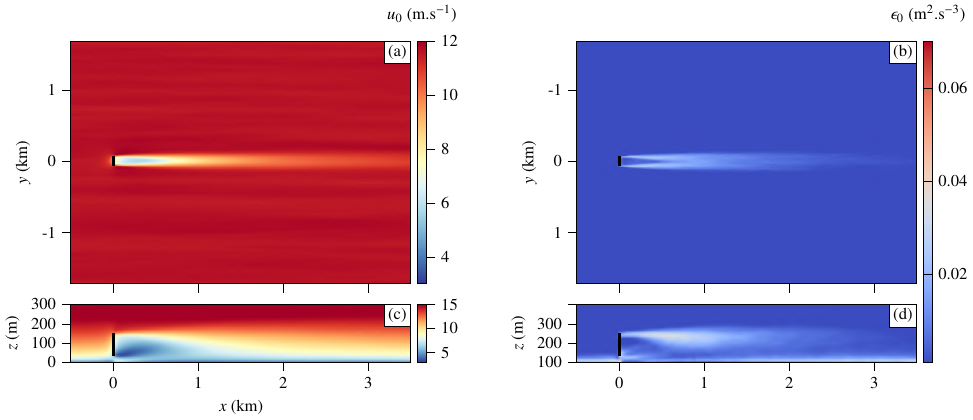}
    \vspace{-0.3cm}
    \caption{ Flow fields for the single turbine layout. \aa and \cc streamwise wind speed $u_0$, \bb and \dd turbulence dissipation rate $\epsilon_0$ for planes at \aa and \bb $z=90~$m (hub height) and at \cc and \dd $y=0$. The rotor position is shown with black lines.}
    \label{f: field_u_1t}
\end{figure*}
The streamwise velocity fields are illustrated in Figs.~\ref{f: field_u_1t}\aa and \cc for an isolated wind turbine.
The figure shows that in these stable conditions, the low turbulence intensity leads to large velocity deficits and long wakes \citep{abkarInfluenceAtmosphericStability2015}.
In addition to wind speed, the acoustic model also uses the turbulence dissipation rate.  
Figs.~\ref{f: field_u_1t}\bb and \dd show that it increases inside the wake. 
Specifically, the increase occurs at the wake edges, where strong shear is present and reflects enhanced turbulent mixing at the wake edges. 
Turbulence dissipation recovers slowly due to the very low turbulence intensity of the incoming flow, consistent with the observations of \citet{abkarInfluenceAtmosphericStability2015}.

\begin{figure*}[htb]
    \footnotesize
    \centering
    \includegraphics{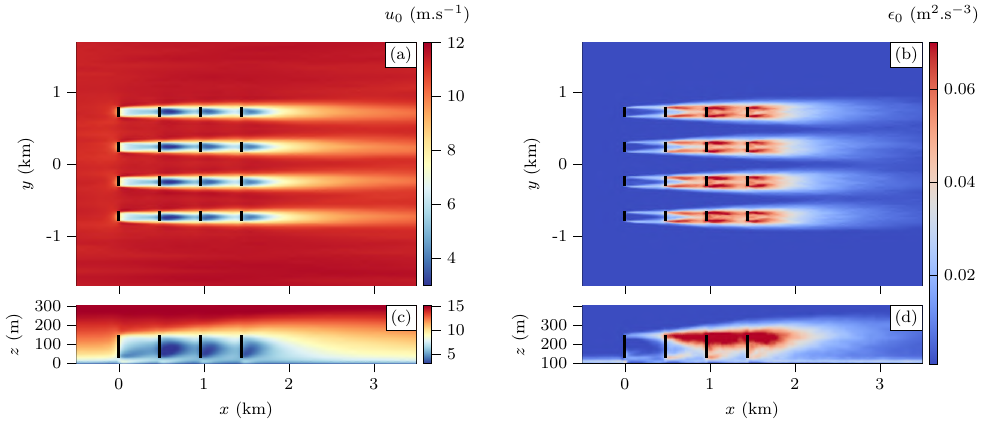}
    \vspace{-0.3cm}
    \caption{ Flow fields for the aligned layout and stable atmospheric condition. \aa and \cc streamwise wind speed $u_0$, \bb and \dd turbulence dissipation rate $\epsilon_0$ for planes at \aa and \bb $z=90~$m (hub height) and at \cc and \dd $y=-720~$m. The rotor positions are shown with black lines.}
    \label{f6: flow_aligned}
\end{figure*}
Figs.~\ref{f6: flow_aligned}(a, c) illustrate the streamwise velocity in the aligned wind farm.
In this case, the second row experiences the largest velocity deficit, while partial wind speed recovery occurs in the third and fourth rows due to energy entrainment from above the wind farm.
In Figs.~\ref{f6: flow_aligned}(b) and (d), the turbulent dissipation rate is plotted for the planes $z=90~$m and $y=-720~$m.
The turbulent dissipation increases inside the wind farm similarly to the results for the single turbine.
The increased mixing at the top of the farm, along with the associated rise in dissipation, explains the wind speed recovery after the second row.
This increase of turbulence within the wind farm may also affect the sound emitted from the turbines.

\begin{figure*}[htb!]
    \footnotesize
    \centering
    \includegraphics{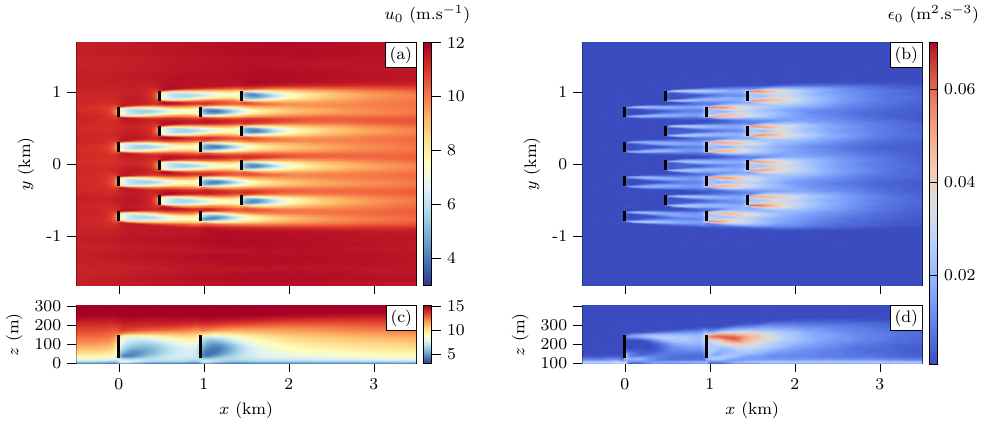}
    \vspace{-0.3cm}
    \caption{ Flow fields for the staggered layout and stable atmospheric condition. \aa and \cc streamwise wind speed $u_0$, \bb and \dd turbulence dissipation rate $\epsilon_0$ for planes at \aa and \bb $z=90~$m (hub height) and at \cc and \dd $y=-720~$m. The rotor positions are shown with black lines.}
    \label{f6: flow_staggered}
\end{figure*}
The flow fields for the staggered case are shown in Fig.~\ref{f6: flow_staggered}.
Due to a larger spacing between turbines, the overall velocity deficit within the farm is reduced in Figs.~\ref{f6: flow_staggered}(a) and (c).
Similarly, the maximum value of the turbulence dissipation rate reached inside the farm is lower than for the aligned case.
This staggered configuration is more representative of actual wind farms than the aligned one, as it reduces wake effects between neighboring turbines.

%%%%%%%%%%%%%%%%%%%%%%%%%%%%%%%%%%%%%%%%%%%%%%%%%%%%%%%%%%%%%%%%%%%%%%%%%%%%%%%
%%%%%%%%%%%%%%%%%%%%%%%%%%%%%%%%%%%%%%%%%%%%%%%%%%%%%%%%%%%%%%%%%%%%%%%%%%%%%%%
\section{Turbine noise emission}
\label{s: source}
\subsection{Input flow}
\begin{figure*}[htbp!]
	\centering
    \footnotesize
    \includegraphics{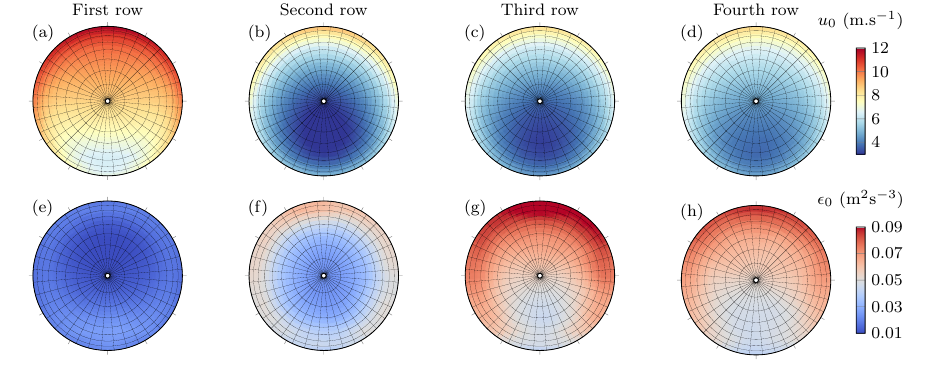}
    \vspace{-0.3cm}
    \caption{Flow input for the source model for the aligned wind farm. Figures \aa to \dd show the wind speed and figures \ee to \hh the turbulent dissipation rate. Columns from left to right indicate results for first to fourth row. 
The rotor discretization for the source model is shown with black lines.}
    \label{f: flow_sample}
\end{figure*}

Fig.~\ref{f: flow_sample} illustrates the mean wind speed and turbulence dissipation rate sampled at the rotor position in each row of the aligned wind farm. 
For conciseness, the flow sampled at the rotor for the staggered wind farm are not presented here. 
The wind velocity drops sharply between the first and second rows, followed by visible recovery in the third and fourth rows. 
Additionally, wind speed varies across the rotor plane.
For all turbines, the ABL shear causes higher wind speeds at the top of the rotor and lower speeds at the bottom. 
Turbine flow blockage also influences wind speed, resulting in slightly higher wind speeds at the rotor edges compared to the center.

The turbulent dissipation rate $\epsilon_0$ influences the TIN and is also depicted in Fig.~\ref{f: flow_sample}.
As previously noted $\epsilon_0$ increases within the wind farm and hence increases from one row to the next.
At the first rotor, we observe a more typical ABL structure, with slightly higher turbulence dissipation at the bottom of the turbine.
For the third and fourth rows the turbulent dissipation rate is higher at the top of the rotor.
This is due to the higher wind speeds, which induce stronger shear and enhanced  wake mixing, leading to a higher turbulence dissipation rate.

\subsection{Sound power levels}
\begin{figure*}[htb]
	\centering
    \footnotesize
    \includegraphics{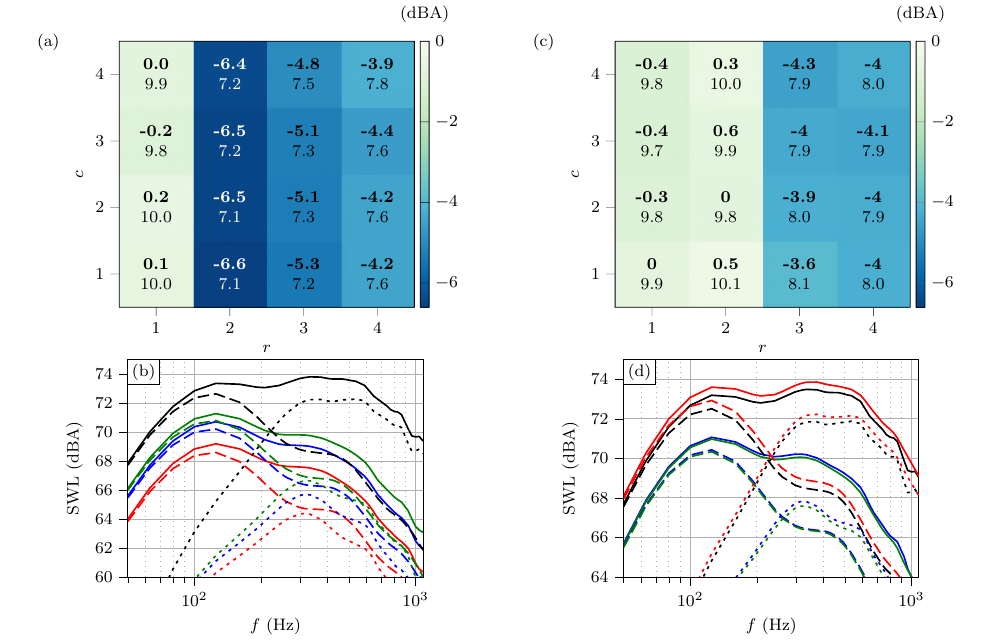}
    \vspace{-0.3cm}
	\caption{OASWL (in bold) relative to the isolated case for each turbine of the \aa aligned and \cc staggered wind farms. The rotational speed $\Omega$ in rpm for each turbine is indicated below. SWL spectra averaged over each column of \bb the aligned and \dd staggered wind farms. Solid lines show the total SWL, dashed and dotted lines show the TIN and TEN contribution, respectively. The first, second, third and fourth rows are shown in black, red, blue and green, respectively.}
	\label{f6: swl}
\end{figure*}
The flow input significantly influences the sound emitted by each wind turbine.
First the model provides an effective overall sound power level of 102.5~dBA for the isolated wind turbine.
Fig.~\ref{f6: swl}\aa displays the overall source power for each turbine in the aligned wind farm relative to the isolated wind turbine.
Turbines in the same row generate similar OASWL within a 0.5~dBA margin.
The first row ($r=1$) has the highest sound power emission, with levels corresponding to the isolated case.
Due to the substantial decrease in wind speed after the first row, which reduces the rotor's rotational speed, the second row ($r=2$) shows the lowest sound power level. 
The OASWL of the third and fourth rows gradually increase as the wind speed recovers within the wind farm.

The frequency content of the produced noise also varies across the wind farm, as illustrated in Fig.~\ref{f6: swl}\bb. 
The sound power is averaged over each row, and the contributions of TIN (dashed lines) and TEN (dotted lines) are shown for each row. 
For the first row (black lines), TIN and TEN contribute evenly, with TIN dominating at low frequencies  and TEN at high frequencies.
In subsequent rows, the contribution of TEN decreases significantly due to the reduced wind speed. 
Although TIN decreases, it does so more slowly than TEN, resulting in a relative stronger contribution and a greater share of low-frequency content. 
Notably, both TIN and TEN contributions increase for the third and fourth rows (blue and green lines) compared to the second row (red line), due to the wind speed recovery within the wind farm. 
For these rows, the TIN contribution (dashed lines) increases slightly faster than the TEN contribution (dotted lines), due to the elevated turbulence dissipation rate, which only influences TIN.

Fig.~\ref{f6: swl}\cc presents the corresponding data for the staggered layout. 
Similarly to the aligned case, the OASWL is very consistent for turbines in the same row.
The two first rows present OASWL very close to that of the isolated turbine with a small increase for the second row. 
This is attributed to the accelerated flow between turbines of the first row (see Fig.~\ref{f6: flow_aligned}\cc), causing slightly higher wind speeds at the second row.  
Furthermore, rows three and four produce considerably more sound than in the aligned case, leading to higher overall noise from the staggered wind farm. 
As shown in Fig.~\ref{f6: flow_staggered}, the flow fields confirm that the last two rows experience similar inflow conditions, which explains their comparable sound production.
As a side note, the staggered wind farm also produces more power than the aligned configuration. We do not analyze this here, as it is already well documented in the literature (see, e.g., \citet{stevensFlowStructureTurbulence2017}).

As expected, the spectra  of the first two rows (black and red lines), shown in Fig.~\ref{f6: swl}\dd, are almost superimposed, as are those of the last two rows (blue and green lines). 
Additionally, like in the aligned layout, low-frequency content becomes more prominent in the last two rows because of the increased TIN contribution.

%%%%%%%%%%%%%%%%%%%%%%%%%%%%%%%%%%%%%%%%%%%%%%%%%%%%%%%%%%%%%%%%%%%%%%%%%%%%%%%
%%%%%%%%%%%%%%%%%%%%%%%%%%%%%%%%%%%%%%%%%%%%%%%%%%%%%%%%%%%%%%%%%%%%%%%%%%%%%%%
\section{$\Delta L$ and propagation effects}
\label{s: deltaL}
In this section the effect of the flow on propagation is studied by considering the $\Delta L$ obtained from the WAPE simulation for a single point source at hub height.

\subsection{Single wake}
\begin{figure}[htbp]
\footnotesize
\centering
\includegraphics{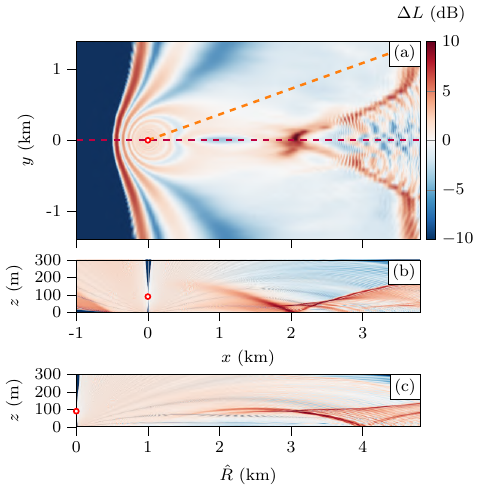}
    \vspace{-0.3cm}
\caption{$\Delta L$ at $f_c=1~$kHz for a source located at hub height of the wind turbine (red dot) for an isolated turbine in the plane (a) $z=2~$m, (b) $y=0~$m (dashed purple line), and (c) $\tau=20$° (dashed orange line). $\hat{R}$ is the radial distance from the turbine.}
\label{f: dl_1T}
\end{figure}
The $\Delta L$ obtained for an isolated turbine at a third-octave center frequency of 1~kHz is shown in Fig.~\ref{f: dl_1T}\aa at $z=2~$m.
Fig.~\ref{f: dl_1T}\bb shows \dl for the streamwise-vertical plane at $y=0$ and Fig.~\ref{f: dl_1T}\cc for a propagation angle $\tau=20$°.
Several features emerge from refraction caused by the mean flow gradient. 
The shadow zone in the upwind direction starts at $x = -200~$m. 
Downwind the wake focusing effect leads to an amplification at the ground at $x=2~$km. 
Starting from this initial focus zone visible in Fig.~\ref{f: dl_1T}\aa, an increase in $\Delta L$  is observed, which moves away from the downwind propagation direction as the receiver’s distance from the source increases, resulting in a "V" shape. 
As the propagation angle increases, the wake’s focusing effect weakens, and its intersection with the ground shifts farther from the source.
This is visible in Fig.~\ref{f: dl_1T}\cc where focusing appears less efficient and reaches the ground at a greater distance ($\hat{R}=4~$km). 
For $|y| > 0.8~$km the focusing appears to follow a different pattern with a $\Delta L$ increase at $x = 3.7~$km independently of the propagation angle. 
This focusing is induced by the ABL velocity gradient itself and does not depend on the wake.

\subsection{Multiple wakes}
\begin{figure*}[htb]
    \footnotesize
    \centering
    \includegraphics{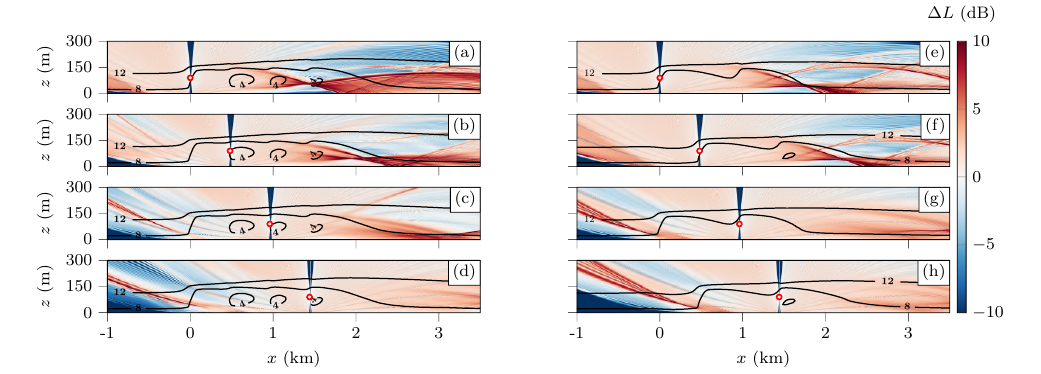}
    \vspace{-0.4cm}
    \caption{$\Delta L$ at $f_c=1~$kHz for a source located at hub height of the wind turbine (red dot) in the first column and for the plane $y=720~$m. The propagation is shown for each row (top to bottom) and for the aligned (left) and staggered (right) cases. The streamwise wind speed contour is shown black lines.}
    \label{f: dl_16T0}
\end{figure*}
The $\Delta L$ fields corresponding to the four turbines positioned in the first column ($c=1$), are depicted for the aligned and staggered cases in Fig.~\ref{f: dl_16T0}.
The \dl from each turbine are shown separately to investigate the propagation effects arising for each row.
The $xz$-planes are located at $y=-720~$m such that they cross the turbine hub.
For the first two rows, wake interactions significantly affect the sound focusing pattern, resulting in two distinct focal regions emerging from the bottom and top of the wake.
Compared to the single turbine case (Fig.~\ref{f: dl_1T}\bb) the focusing is much stronger due to a channeling effect of the wind farm. 
This focusing reaches the ground slightly closer to the source for the aligned case (at $x=1.8~$km) compared to the staggered case (around $x=2.1~$km).
However, overall, the difference in flow patterns has a limited influence on the sound propagation from the first turbine row.
The same holds for the second row.
Again, stronger focusing is observed compared to the single-turbine case, with the sound reaching the ground at approximately $x = 2.5~$km for both the staggered and aligned configurations, which is similar as the first row. 

In contrast, for the turbines in the last two rows, focusing is less efficient than for the single turbine case.
This can be explained as the wind speed gradients behind the wind farm are smaller than those created by a single turbine.
Upwind of the first row the shadow zone appears to be identical to the single turbine case in Fig.~\ref{f: dl_1T}.
For the subsequent rows, due to the wakes of the upstream turbines, the shadow zone is pushed back towards the front of the wind farm.

\begin{figure*}[htb]
    \footnotesize
    \centering
    \includegraphics{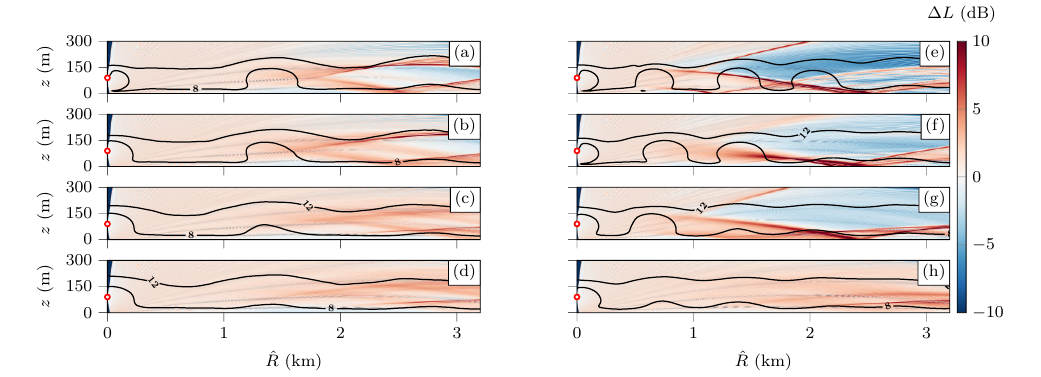}
    \vspace{-0.4cm}
    \caption{$\Delta L$ at $f_c=1~$kHz for a source located at hub height of the wind turbine (red dot) in the first column and for the plane $\tau=20$°. The propagation is shown for each row (top to bottom) and for the aligned (left) and staggered (right) cases. $\hat{R}$ is the radial distance from the turbine. The wind speed projected in the propagation plane is shown with the black contour lines.}
    \label{f: dl_16T20}
\end{figure*}
The wind farm flow influences sound propagation in all directions.
To showcase this, the $\Delta L$ is plotted for the propagation plane $\tau=20$° in Fig.~\ref{f: dl_16T20}. 
For this propagation direction there is a clear difference between the single turbine (Fig.~\ref{f: dl_1T}\cc), the aligned and the staggered wind farm. 
The presence of other wakes in this direction creates another focusing for the first two turbines in the aligned case, see Figs.~\ref{f: dl_16T20}(a, b). 
It reaches the ground about 2.5~km downwind of the wind turbine.
For the turbines of the third and fourth rows the velocity gradients are not strong enough to create focusing,  see Figs.~\ref{f: dl_16T20}(c, d).
In the staggered layout, a strong focusing is observed for the first three rows, Figs.~\ref{f: dl_16T20}(e-g).
This is explained by observing the projected flow fields, see the contour lines.
Due to the staggered layout,  this propagation plane crosses more wakes in the staggered case than in the aligned case.
This induces multiple sound wave focusing zones, resulting in a strong increase in $\Delta L$ at the ground at $R = 2.5~$km.
From this observation we can expect a higher SPL and AM in these propagation directions for $\tau<90$° compared to the single turbine or aligned wind farm.

%%%%%%%%%%%%%%%%%%%%%%%%%%%%%%%%%%%%%%%%%%%%%%%%%%%%%%%%%%%%%%%%%%%%%%%%%%%%%%%
%%%%%%%%%%%%%%%%%%%%%%%%%%%%%%%%%%%%%%%%%%%%%%%%%%%%%%%%%%%%%%%%%%%%%%%%%%%%%%%
\section{OASPL and AM}
\label{s: oaspl}
\subsection{Single turbines}
\begin{figure}[htb]
    \footnotesize
    \centering
    \includegraphics{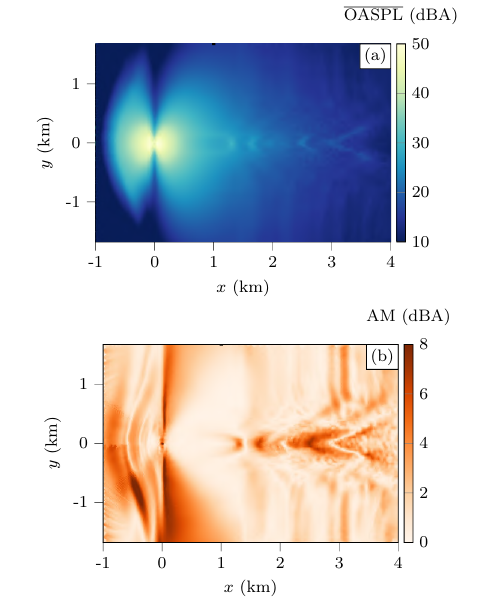}
    \vspace{-0.3cm}
    \caption{(a) \OASPL, (b) AM at $z=2~$m for an isolated turbine.}
    \label{f: oaspl_top_1T}
\end{figure}

Fig.~\ref{f: oaspl_top_1T} shows the $\overline{\rm OASPL}$ and the AM at $z=2~$m for the isolated wind turbine.
This baseline case presents different features characteristic of wind turbine noise propagation. 
The dipolar shape induced by the TIN and TEN directivity is visible in Fig.~\ref{f: oaspl_top_1T}\aa, displaying the characteristic cross-wind extinction zone. 
A shadow region is also present upwind for $x < -300~$m.
It is created by the negative effective sound speed gradient in the upwind propagation direction. 
Finally, some focusing patterns are visible downwind, which can be attributed to the wake and ABL shear. 
The "V" shape that was already visible in the $\Delta L$  fields (see Sec.~\ref{s: deltaL}) is still present.

As previously stated, AM is essential in assessing perception of wind turbine noise. 
Fig.~\ref{f: oaspl_top_1T}\bb shows the AM measured at 2~m height, highlighting cross-wind AM caused by the rotating blades moving toward and away from the receiver. 
Consistent with \citet{oerlemansPredictionWindTurbine2009}, we find that there is an asymmetry in AM cross-wind, the higher AM values are observed for the negative $y$ which corresponds to the observer close to the downward moving blade. 
AM also increases in the upwind direction due to the negative wind speed gradient, similarly to the observation of \citet{mascarenhasPropagationEffectsSynthesis2023}. 
The change in source height due to blade rotation alters the focusing pattern, leading to AM at the ground, visible downwind for $x>1.5~$km.
This increase in AM directly induced by the wake effect is consistent with the work of other authors \citep{barlasEffectsWindTurbine2017,heimannSoundPropagationWake2018}.

\begin{figure*}[htb]
    \centering
    \footnotesize
    \includegraphics{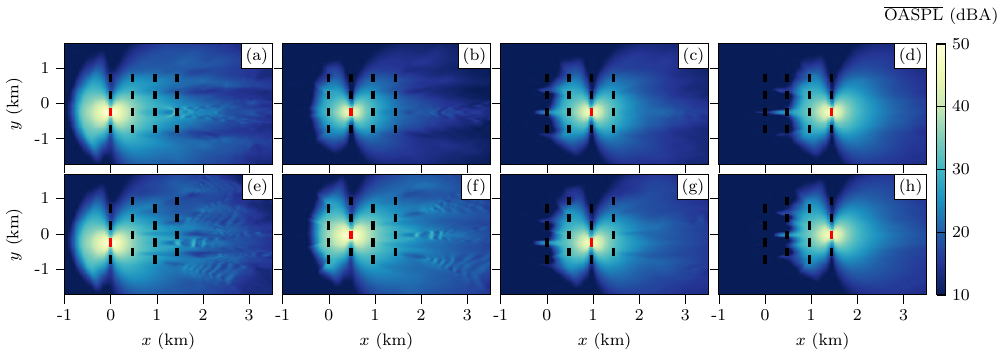}
    \vspace{-0.4cm}
    \caption{Contribution of a single turbine (in red) to the \OASPL at $z=2~$m,  Figs.~\aa to \dd correspond to the aligned layout and Figs.~\ee to \hh to the staggered layout. Wind turbine rotors are indicated with black lines.}
    \label{f: oaspl_compar}
\end{figure*}

To further study the impact of multiple wakes on sound propagation, Fig.~\ref{f: oaspl_compar} shows the averaged OASPL at 2~m for one turbine in each row of the aligned and the staggered cases.
The effect of the additional wakes on sound propagation is visible both for the staggered and aligned layouts. 
For the aligned case, Figs.~\ref{f: oaspl_compar}(a-d), areas of noise amplification and noise reduction are present at almost constant $y$ values.
The wakes of the first, third and forth columns are responsible for these patterns by focusing the sound wave in some specific regions.
These patterns are mostly visible for the two first turbines and are less present for the last turbine.
This corroborates with the $\Delta L$ results shown in the previous section. 
For the staggered case, see Figs.~\ref{f: oaspl_compar}(e-h), the focusing pattern appears less symmetrical which can be explained by the turbine arrangement, which leads to more complex wake physics in the wind farm.
Focusing also appears more efficient for the first two rows, which again is in agreement with observations on $\Delta L$. 

\begin{figure*}[htb]
    \centering
    \footnotesize
    \includegraphics{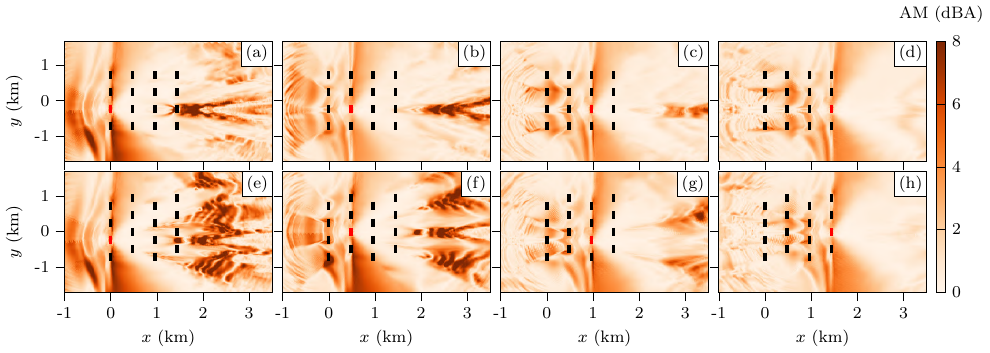}
    \vspace{-0.4cm}
    \caption{Amplitude modulation generated by a single turbine (in red) at $z=2~$m, Figs.~\aa to \dd correspond to the aligned layout and Figs.~\ee to \hh to the staggered layout. Wind turbine rotors are indicated with black lines.}
    \label{f: am_compar}
\end{figure*}

The wind farm flow also modifies the AM pattern as demonstrated in Fig.~\ref{f: am_compar}.
% downwind 
In the aligned case the AM increases strongly in the downwind direction for the first two rows, see Figs.~\ref{f: am_compar}\aa and \bb. 
Due to the wake superposition, focusing is more efficient which leads to significant variations in OASPL as the blade rotates. 
For the last turbine, Fig.~\ref{f: am_compar}\dd, no AM increase is visible downwind. 
As shown in the previous section, focusing is less efficient for the last row, leading to reduced AM.
% cross wind
The first row in Fig.~\ref{f: am_compar}\aa presents similar crosswind AM as the isolated turbine, see Fig.~\ref{f: oaspl_top_1T}\bb.
Crosswind AM is reduced for all the subsequent rows due to reduced rotational speed and hence less convective amplification effect.
% upwind
Finally, the first row shows upwind AM similar to the single-turbine case, as the flow is similar in both cases.
For the subsequent rows the upwind AM pattern is modified by the wakes upwind of the turbine.
However, the AM increase is mostly present inside the wind farm and hence is not particularly relevant. 

For the staggered layout, Figs.~\ref{f: am_compar}\ee and \ff show a downwind increase in AM for the first two rows. 
Additionally, significant AM increase is also visible for other propagation angles ($|\tau|<90$).
This supports the $\Delta L$ results which shows enhanced focusing for $\tau=20$° compared to the aligned case.
Figs.~\ref{f: am_compar}\gg and \hh display no direct downwind AM for the third and fourth rows, but an increase in AM for the third row is visible for $\tau=20$° and $\tau=-20$°.
This is again in agreement with our $\Delta L$ results.
% crosswind
Crosswind AM is similar for the first two rows, as shown in Figs.~\ref{f: am_compar}\ee and \ff, which is expected given their nearly identical SWL.
The crosswind AM for the last two rows are lower, see Figs.~\ref{f: am_compar}\gg and \hh. 
% upwind 
The upwind AM for the first row is similar to that in the aligned wind farm  and the single-turbine case.
In contrast, the upwind AM for turbines in rows 2-4 are influence by the flow within the wind farm.

%%%%%%%%%%%%%%%%%%%%%%%%%%%%%%%%%%%%%%%%%%%%%%%%%%%%%%%%%%%%%%%%%%%%%%%%%%%%%%%
\subsection{Full wind farm}
\begin{figure*}[htb]
    \centering
    \footnotesize
    \includegraphics{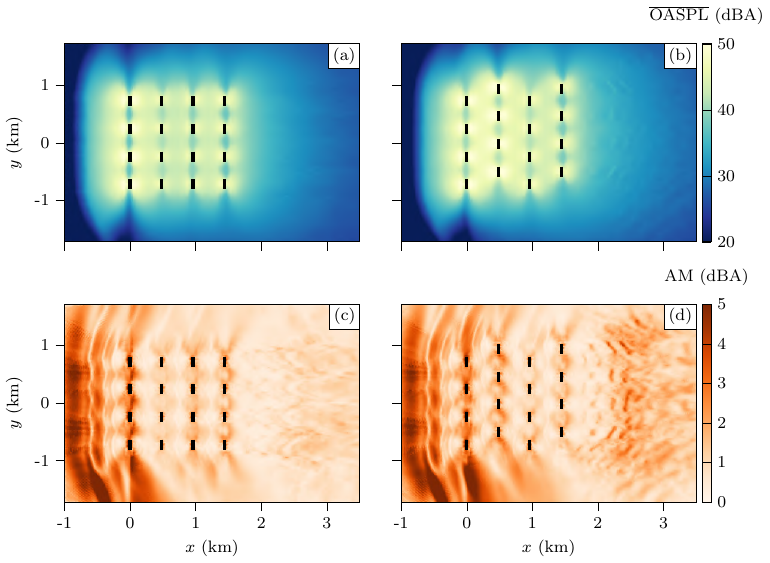}
    \vspace{-0.3cm}
    \caption{\aa, \bb \OASPL and \cc, \dd AM for the \aa, \cc aligned and \bb, \dd staggered layouts at $z=2~$m. Wind turbine rotors are indicated with black lines.}
    \label{f: oaspl_top_16T}
\end{figure*}

Figs.~\ref{f: oaspl_top_16T}\aa and \bb display the \OASPL fields at 2~m, considering the contribution of 16 wind turbines. 
In the aligned case, each turbine's distinct contributions are visible, with the first row emitting higher sound levels. 
Upwind, a shadow zone starts at $x=-500~$m, while downwind, the focusing patterns seen in the single turbine case are absent, as expected due to the weak focusing from the last row's wakes and the averaging effect of multiple turbines.
Cross-wind extinction zones ($y < -720~$m and $y > 720~$m) are significantly attenuated for the second ($x=480~$m), third ($x=960~$m), and fourth ($x=1.44~$km) rows due to upwind turbines' contributions. 
The first row ($x$=0) shows two cross-wind extinction zones, explained by the low sound power level from the second row's turbines (Sec.~\ref{s: source}) and upward refraction reducing noise contribution of the downwind turbines.
The staggered layout \OASPL field differs from the aligned layout, with higher \OASPL near the wind turbine for the second row and slight focusing patterns downwind at around $x=2.5~$km.

Fig.~\ref{f: oaspl_top_16T}\cc shows the AM field for the aligned layout.
Compared to the isolated case only crosswind AM for the first row and upwind AM are present in the aligned case.
The downwind AM generated by individual turbine wakes, which appeared more pronounced when viewed separately, is not visible in the total field.
This is explained by a smoothing of the amplification due to the contribution of multiple turbines.
As previously stated the AM cross wind for the first row is still visible because the contribution of the subsequent rows is very low in this region.
Similarly, the upwind AM has a comparable amplitude to that of the isolated turbine.

For the staggered layout, Fig.~\ref{f: oaspl_top_16T}\dd, there is a significant increase in AM downwind ($2 < x < 2.5~$km) compared to the aligned case.
In this case, AM increases for turbines in the first and second rows across multiple propagation directions. 
Additionally, the last rows emit more noise, contributing to higher AM levels downstream.
The location of increased AM downwind in Fig.~\ref{f: oaspl_top_16T}\dd aligns with the AM rise observed for the first and second rows in Figs.~\ref{f: am_compar}\ee and \gg.
This leads to think that the first and second rows are responsible for the increased AM downwind. 
Upwind and crosswind, AM are similar to that of the aligned case.

\begin{figure}[htb]
	\centering
    \footnotesize
    \includegraphics{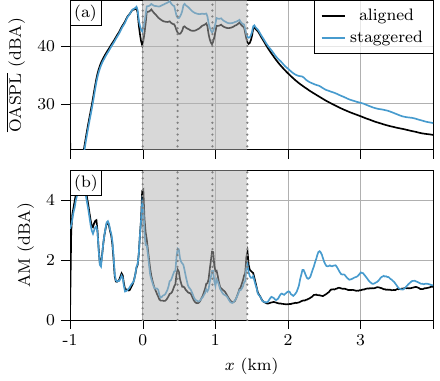}
	\vspace{-0.2cm}
	\caption{Spatially averaged (a) \OASPL and (b) AM  for the aligned and staggered layouts for receivers at $z=2~$m. The \OASPL and AM values are averaged over $|y|<720~$m. Wind turbines' positions are indicated with gray dotted lines, the gray area represents the positions inside the wind farm.}
	\label{f: oaspl_line}
\end{figure}

%% Line plot 
To further investigate the differences between the aligned and staggered layouts,
Fig.~\ref{f: oaspl_line} compares the \OASPL and AM, averaged over $|y|<720~$m, for the aligned and staggered layouts.
Upwind both configurations show similar results.
However, downwind of the wind farm  ($x > 2~$km), the staggered case demonstrates a significant increase of $3~$dBA in \OASPL due to the increase in the sound emitted by each turbine. 
Another significant feature is the increase in AM at around $x=2.5~$km, see Fig.~\ref{f: oaspl_line}\bb. 
The maximum downwind AM is 1~dBA for the aligned layout and 2~dBA for the staggered layout, which is usually considered as a threshold for perceiving AM \citep{nguyenLongtermQuantificationCharacterisation2021}. 

\section{Wind farm flow effect on sound propagation}
To further assess the effect of the wind farm flow physics on sound propagation, the \OASPL from the wind farm is computed using the $\Delta L$ computed for the single turbine case.
This enables the computation of wind farm sound propagation using realistic sound power levels, but without sound refraction induced by the wake interaction.
This artificial case is then subtracted from the real wind farm case to obtain
\begin{equation}
    \Delta {\overline{\rm OASPL}}  = {\overline{\rm OASPL}}_{16T}^{\rm tot} - {\overline{\rm OASPL}}_{1T}^{\rm tot}
\end{equation}
where OASPL$_{16T}^{\rm tot}$ and OASPL$_{1T}^{\rm tot}$ are the total fields obtained by integrating over all frequencies and by logarithmically summing the contribution from each turbine $(t_r^c)$, obtained with
\begin{equation}
    \begin{aligned}
        {\rm SPL}_{16T}(t_r^c) & = {\rm SWL}(t_r^c) + \Delta L (t_r^c) - \alpha R  -10\log_{10}(4\pi R^2)\\
        {\rm SPL}_{1T}(t_r^c) &= {\rm SWL}(t_r^c) + \Delta L_{1T} - \alpha R - 10\log_{10}(4\pi R^2)
\end{aligned}
\end{equation}
where SWL$(t_r^c)$ is the sound power obtained for each turbine, \dl$(t_r^c)$ is the sound pressure level relative to the free field of each turbine in the wind farm, and $\Delta L_{1T}$ is the sound pressure level relative to the free field corresponding to the single turbine case.
Note that the actual procedure is a bit more complex because we use different \dl according to the blade position.
But the general idea is that between both computations only the propagation effect is modified.

\begin{figure}[htb]
    \centering
    \footnotesize
    \includegraphics{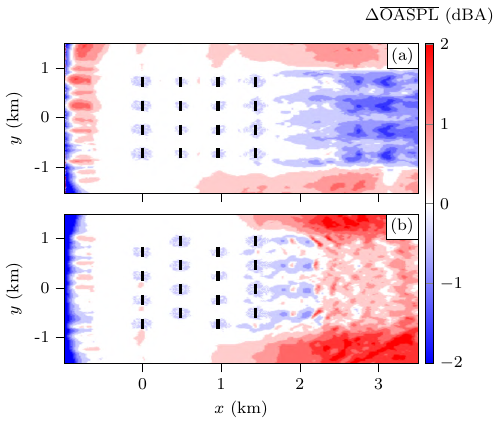}
    \caption{In the \aa aligned and \bb staggered cases, difference between the \OASPL obtained with the full wind farm flows and without accounting for flow interactions between turbines.}
    \label{f: oaspl_diff_T16}
\end{figure}
Fig.~\ref{f: oaspl_diff_T16} shows the corresponding impact of realistic wind farm flow physics on sound propagation at 2~m height.
Inside the wind farms, the difference is close to 0 indicating a minimal impact of the flow.
However, downstream, significant differences are observed.
Directly downwind of the aligned wind farm, see Fig.~\ref{f: oaspl_diff_T16}\aa, for $|y|<720~$m, the \OASPL is between $1~$dBA and $2~$dBA lower when accounting for the realistic wind farm flow. 
On the contrary, outside the direct downwind direction ($|y|>720~$m) an increase is observed due to the combined wake effect.
As shown in the previous section, multiple wakes can create focusing outside the downwind direction, which is not present if only a single wake is considered.
Upwind there is an increase when considering the realistic wind farm flow. 
Note that this increase inside the shadow zone is not significant, and that it can be explained by the fact that the shadow zones of the second, third and fourth rows are pushed back due to the auxiliary wakes.  

The effect of the staggered wind farm flow on propagation is shown in  Fig.~\ref{f: oaspl_diff_T16}\bb. 
Here an even stronger effect than for the aligned case is visible.
Note that only propagation effects can cause differences in \OASPL.
Consequently, the staggered layout has a strong impact on downwind propagation, resulting in a general increase in sound levels.
In comparison, for the aligned case, there was a decrease directly downwind of the wind farm and an increase outside the direct downwind direction.
This confirms that the staggered case is indeed a worst case scenario for acoustics due to both an increase in sound emitted by the turbines and propagation effects that tend to increase the \OASPL downwind.

\section{Beating effect}
\begin{figure*}[htb]
    \centering
    \footnotesize
    \includegraphics{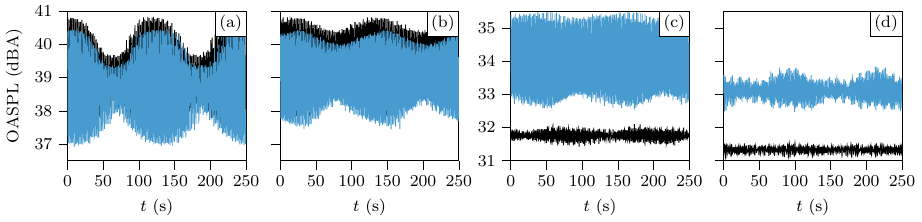}
	\vspace{-0.4cm}
\caption{OASPL time signal at four receiver locations in the aligned (black) and staggered (blue) wind farms. Figures (a) $(x,y)=(-500,0)~$m and (b) $(x,y)=(-500,140)~$m correspond to upwind receivers and Figures (c) $(x,y)=(2000,0)~$m and (d) $(x,y)=(2000,140)~$m to downwind receivers.}
\label{f: signals}
\end{figure*}

Finally, it is important to note that turbines in a wind farm rotate at different speeds. 
As a result, signals with slightly different periods add up and a beating effect can be observed at certain locations, as described by \citet{vandenbergBeatGettingStronger2005}. 
This effect occurs when turbine contributions have similar magnitudes and their rotational speeds are close, producing a signal modulated by a low-frequency envelope relative to the blade-passing frequency. 
Fig.~\ref{f: signals} shows the OASPL signals at four receiver locations for the aligned and staggered wind farm layouts. 

Figs.~\ref{f: signals}(a, b) correspond to upwind receivers, while Figs.~\ref{f: signals}(c, d)  show downwind positions. 
Beating is evident at both upwind locations in both layouts, with AM levels ranging from approximately 4~dBA to below 2~dBA, exceeding the perception threshold reported by \citet{nguyenLongtermQuantificationCharacterisation2021}. 
Downwind of the aligned wind farm, AM levels are very low. 
In contrast, behind the staggered wind farm, a weaker beating effect remains visible. 
The presence of multiple turbines likely smooths the signal, reducing the modulation depth. This beating adds an extra layer of variability, as AM itself is intermittent and sometimes perceptible. 
Intermittent sounds tend to be more noticeable than continuous ones, especially when they stand out from background noise \citep{leeAnnoyanceCausedAmplitude2011}.

%%%%%%%%%%%%%%%%%%%%%%%%%%%%%%%%%%%%%%%%%%%%%%%%%%%%%%%%%%%%%%%%%%%%%%%%%%%%%%%
%%%%%%%%%%%%%%%%%%%%%%%%%%%%%%%%%%%%%%%%%%%%%%%%%%%%%%%%%%%%%%%%%%%%%%%%%%%%%%%
\section{Discussion and conclusion}
\label{s: conclusion}

This study investigated the influence of wind farm layout and flow physics on noise generation and propagation, using large-eddy simulations coupled with acoustic modeling. 
Two wind farm configurations, aligned and staggered, were analyzed, with an isolated turbine case serving as a baseline. 
The results for overall sound pressure level and amplitude modulation in the reference case aligned with earlier findings by \citet{barlasVariabilityWindTurbine2018} and \citet{heimann3DsimulationSoundPropagation2018}.

The wind farm layout has a clear effect on both the overall noise levels and the way sound propagates downwind, consistent with earlier work by \citet{sunDevelopmentEfficientNumerical2018}. 
In aligned layouts, strong wake effects lead to more variation in noise production across turbine rows. In the first row, turbulent inflow noise (TIN) and trailing edge noise (TEN)  contribute about equally, with TIN dominating at low frequencies and TEN at higher frequencies. 
Further downstream, TEN decreases due to reduced wind speeds, while TIN remains relatively high because of increased turbulence dissipation.
Staggered layouts result in higher overall noise levels, as turbines experience stronger inflow and less variation in turbulence dissipation rate inside the farm.

The physics of wind farm flow also significantly affect sound propagation with additional focusing occurring downwind compared to an isolated case due to the wakes of the other turbines in the farm. 
This aligns with the findings of \citet{sunDevelopmentEfficientNumerical2018} who showed similar effects when considering wake superposition.
\citet{nyborgOptimizationWindFarm2023} also demonstrated that wind farm flow has an influence on the noise level and hence should be accounted for when implementing curtailment plans.
The staggered layout presents enhanced focusing downwind leading to increase noise level and amplitude modulation. 
We also showed that these effects where not reproducible without accounting for the multiple wake interaction hence underscoring the importance of capturing full wind farm flow physics when predicting acoustic impact.

This study used time-averaged flow fields to identify persistent effects of wind farm configuration.
Future work should incorporate unsteady simulations and validation with field data, such as measurements by \citet{mascarenhasPropagationEffectsSynthesis2023} and \citet{koneckeNewBaseWind2023}, to assess model accuracy under realistic atmospheric conditions. 
Overall, the findings support the need to include realistic flow fields in acoustic models and provide a foundation for more accurate noise impact assessments, which are essential for wind farm design, permitting, and public acceptance.

\section*{Acknowledgements}
This work was performed within the framework of the LABEX CeLyA (ANR-10-LABX-0060) of Université de Lyon, within the program “Investissements d’Avenir” (ANR-16-IDEX-0005) operated by the French National Research Agency (ANR). 
This project has received funding from the European Research Council under the European Union’s Horizon Europe program (Grant No. 101124815).
It was granted access to the HPC resources of PMCS2I (Pôle de Modélisation et de Calcul en Sciences de l’Ingénieur et de l’Information) of Ecole Centrale de Lyon. 
This work was supported by the Franco-Dutch Hubert Curien partnership (Van Gogh Programme No. 49310UM). 
For the purpose of Open Access, a CC-BY public copyright license has been applied by the authors to the present document and will be applied to all subsequent versions up to the Author Accepted Manuscript arising from this submission.

\section*{Credit authorship contribution}
\textbf{Jules Colas:} conceptualization, investigation, methodology, validation, vizualisation, writing - original draft.  
\textbf{Ariane Emmanuelli, Didier Dragna, Richard Stevens:} conceptualization, methodology, supervision, writing - review \& editing.

\section*{Data availability}
Reported data can be made available to the reader upon request to
the authors.

\section*{Declaration of competing interest}
The authors declare that they have no known competing financial
interests or personal relationships that could have appeared to influence
the work reported in this paper.

% \input{abreviate.bbl}
% \bibliography{zotero3}
% \bibliography{staticZoteroBib}

\begin{thebibliography}{34}
\expandafter\ifx\csname natexlab\endcsname\relax\def\natexlab#1{#1}\fi
\providecommand{\url}[1]{\texttt{#1}}
\providecommand{\href}[2]{#2}
\providecommand{\path}[1]{#1}
\providecommand{\DOIprefix}{doi:}
\providecommand{\ArXivprefix}{arXiv:}
\providecommand{\URLprefix}{URL: }
\providecommand{\Pubmedprefix}{pmid:}
\providecommand{\doi}[1]{\href{http://dx.doi.org/#1}{\path{#1}}}
\providecommand{\Pubmed}[1]{\href{pmid:#1}{\path{#1}}}
\providecommand{\bibinfo}[2]{#2}
\ifx\xfnm\relax \def\xfnm[#1]{\unskip,\space#1}\fi
%Type = Techreport
\bibitem[{Costanzo and Brindley(2024)}]{costanzoWindEnergyEurope2024}
\bibinfo{author}{G.~Costanzo}, \bibinfo{author}{G.~Brindley},
  \bibinfo{title}{Wind Energy in {{Europe}}: 2023 Statistics and the Outlook
  for 2024-2030}, \bibinfo{type}{Technical Report}, Wind Europe,
  \bibinfo{year}{2024}.
%Type = Article
\bibitem[{Barlas et~al.(2017)Barlas, Zhu, Shen, Kelly, and
  Andersen}]{barlasEffectsWindTurbine2017}
\bibinfo{author}{E.~Barlas}, \bibinfo{author}{W.~J. Zhu},
  \bibinfo{author}{W.~Z. Shen}, \bibinfo{author}{M.~Kelly},
  \bibinfo{author}{S.~J. Andersen},
\newblock \bibinfo{title}{Effects of wind turbine wake on atmospheric sound
  propagation},
\newblock \bibinfo{journal}{Appl. Acoust.} \bibinfo{volume}{122}
  (\bibinfo{year}{2017}) \bibinfo{pages}{51--61}.
  \DOIprefix\doi{10.1016/j.apacoust.2017.02.010}.
%Type = Article
\bibitem[{Heimann and
  Englberger(2018)}]{heimann3DsimulationSoundPropagation2018}
\bibinfo{author}{D.~Heimann}, \bibinfo{author}{A.~Englberger},
\newblock \bibinfo{title}{{{3D-simulation}} of sound propagation through the
  wake of a wind turbine: {{Impact}} of the diurnal variability},
\newblock \bibinfo{journal}{Appl. Acoust.} \bibinfo{volume}{141}
  (\bibinfo{year}{2018}) \bibinfo{pages}{393--402}.
  \DOIprefix\doi{10.1016/j.apacoust.2018.06.005}.
%Type = Article
\bibitem[{Kayser et~al.(2020)Kayser, Cott{\'e}, Ecoti{\`e}re, and
  Gauvreau}]{kayserEnvironmentalParametersSensitivity2020}
\bibinfo{author}{B.~Kayser}, \bibinfo{author}{B.~Cott{\'e}},
  \bibinfo{author}{D.~Ecoti{\`e}re}, \bibinfo{author}{B.~Gauvreau},
\newblock \bibinfo{title}{Environmental parameters sensitivity analysis for the
  modeling of wind turbine noise in downwind conditions},
\newblock \bibinfo{journal}{J. Acoust. Soc. Am.}
  \bibinfo{volume}{148} (\bibinfo{year}{2020}) \bibinfo{pages}{3623--3632}.
  \DOIprefix\doi{10.1121/10.0002872}.
%Type = Article
\bibitem[{Bommidala et~al.(2025)Bommidala, Colas, Emmanuelli, Dragna, Khodr,
  Cott{\'e}, and Stevens}]{bommidalaThreedimensionalEffectsWake2025}
\bibinfo{author}{H.~Bommidala}, \bibinfo{author}{J.~Colas},
  \bibinfo{author}{A.~Emmanuelli}, \bibinfo{author}{D.~Dragna},
  \bibinfo{author}{C.~Khodr}, \bibinfo{author}{B.~Cott{\'e}},
  \bibinfo{author}{R.J.A.M. Stevens},
\newblock \bibinfo{title}{Three-dimensional effects of the wake on wind turbine
  sound propagation using parabolic equation},
\newblock \bibinfo{journal}{J. Sound Vib.}
  \bibinfo{volume}{608} (\bibinfo{year}{2025}) \bibinfo{pages}{119036}.
  \DOIprefix\doi{10.1016/j.jsv.2025.119036}.
%Type = Article
\bibitem[{{Port{\'e}-Agel} et~al.(2020){Port{\'e}-Agel}, Bastankhah, and
  Shamsoddin}]{porte-agelWindTurbineWindFarmFlows2020}
\bibinfo{author}{F.~{Port{\'e}-Agel}}, \bibinfo{author}{M.~Bastankhah},
  \bibinfo{author}{S.~Shamsoddin},
\newblock \bibinfo{title}{Wind-{{Turbine}} and {{Wind-Farm Flows}}: {{A
  Review}}},
\newblock \bibinfo{journal}{Bound.-Layer Meteorol.} \bibinfo{volume}{174}
  (\bibinfo{year}{2020}) \bibinfo{pages}{1--59}.
  \DOIprefix\doi{10.1007/s10546-019-00473-0}.
%Type = Article
\bibitem[{Stevens and Meneveau(2017)}]{stevensFlowStructureTurbulence2017}
\bibinfo{author}{R.J.A.M. Stevens}, \bibinfo{author}{C.~Meneveau},
\newblock \bibinfo{title}{Flow {{Structure}} and {{Turbulence}} in {{Wind
  Farms}}},
\newblock \bibinfo{journal}{Annu. Rev. Fluid Mech.}
  \bibinfo{volume}{49} (\bibinfo{year}{2017}) \bibinfo{pages}{311--339}.
  \DOIprefix\doi{10.1146/annurev-fluid-010816-060206}.
%Type = Article
\bibitem[{Shen et~al.(2019)Shen, Zhu, Barlas, and
  Li}]{shenAdvancedFlowNoise2019}
\bibinfo{author}{W.~Z. Shen}, \bibinfo{author}{W.~J. Zhu},
  \bibinfo{author}{E.~Barlas}, \bibinfo{author}{Y.~Li},
\newblock \bibinfo{title}{Advanced flow and noise simulation method for wind
  farm assessment in complex terrain},
\newblock \bibinfo{journal}{Renew. Energy} \bibinfo{volume}{143}
  (\bibinfo{year}{2019}) \bibinfo{pages}{1812--1825}.
  \DOIprefix\doi{10.1016/j.renene.2019.05.140}.
%Type = Article
\bibitem[{Sun et~al.(2018)Sun, Zhu, Shen, Barlas, S{\o}rensen, Cao, and
  Yang}]{sunDevelopmentEfficientNumerical2018}
\bibinfo{author}{Z.~Sun}, \bibinfo{author}{W.~Zhu}, \bibinfo{author}{W.~Shen},
  \bibinfo{author}{E.~Barlas}, \bibinfo{author}{J.~S{\o}rensen},
  \bibinfo{author}{J.~Cao}, \bibinfo{author}{H.~Yang},
\newblock \bibinfo{title}{Development of an {{Efficient Numerical Method}} for
  {{Wind Turbine Flow}}, {{Sound Generation}}, and {{Propagation}} under
  {{Multi-Wake Conditions}}},
\newblock \bibinfo{journal}{Appl. Sci.} \bibinfo{volume}{9}
  (\bibinfo{year}{2018}) \bibinfo{pages}{100}.
  \DOIprefix\doi{10.3390/app9010100}.
%Type = Article
\bibitem[{Nyborg et~al.(2023)Nyborg, Fischer, R{\'e}thor{\'e}, and
  Feng}]{nyborgOptimizationWindFarm2023}
\bibinfo{author}{C.~M. Nyborg}, \bibinfo{author}{A.~Fischer},
  \bibinfo{author}{P.-E. R{\'e}thor{\'e}}, \bibinfo{author}{J.~Feng},
\newblock \bibinfo{title}{Optimization of wind farm operation with a noise
  constraint},
\newblock \bibinfo{journal}{Wind Energy Science} \bibinfo{volume}{8}
  (\bibinfo{year}{2023}) \bibinfo{pages}{255--276}.
  \DOIprefix\doi{10.5194/wes-8-255-2023}.
%Type = Article
\bibitem[{Colas et~al.(2023)Colas, Emmanuelli, Dragna, {Blanc-Benon},
  Cott{\'e}, and Stevens}]{colasWindTurbineSound2023}
\bibinfo{author}{J.~Colas}, \bibinfo{author}{A.~Emmanuelli},
  \bibinfo{author}{D.~Dragna}, \bibinfo{author}{P.~{Blanc-Benon}},
  \bibinfo{author}{B.~Cott{\'e}}, \bibinfo{author}{R.J.A.M. Stevens},
\newblock \bibinfo{title}{Wind turbine sound propagation: {{Comparison}} of a
  linearized {{Euler}} equations model with parabolic equation methods},
\newblock \bibinfo{journal}{J. Acoust. Soc. Am.}
  \bibinfo{volume}{154} (\bibinfo{year}{2023}) \bibinfo{pages}{1413--1426}.
  \DOIprefix\doi{10.1121/10.0020834}.
%Type = Article
\bibitem[{Gadde et~al.(2021)Gadde, Stieren, and
  Stevens}]{gaddeLargeEddySimulationsStratified2021}
\bibinfo{author}{S.~N. Gadde}, \bibinfo{author}{A.~Stieren},
  \bibinfo{author}{R.J.A.M. Stevens},
\newblock \bibinfo{title}{Large-{{Eddy Simulations}} of {{Stratified
  Atmospheric Boundary Layers}}: {{Comparison}} of {{Different Subgrid
  Models}}},
\newblock \bibinfo{journal}{Bound.-Layer Meteorol.} \bibinfo{volume}{178}
  (\bibinfo{year}{2021}) \bibinfo{pages}{363--382}.
  \DOIprefix\doi{10.1007/s10546-020-00570-5}.
%Type = Article
\bibitem[{Gadde and Stevens(2019)}]{gaddeEffectCoriolisForce2019}
\bibinfo{author}{S.~N. Gadde}, \bibinfo{author}{R.J.A.M. Stevens},
\newblock \bibinfo{title}{Effect of {{Coriolis}} force on a wind farm wake},
\newblock \bibinfo{journal}{J. Phys.: Conf. Ser.}
  \bibinfo{volume}{1256} (\bibinfo{year}{2019}) \bibinfo{pages}{012026}.
  \DOIprefix\doi{10.1088/1742-6596/1256/1/012026}.
%Type = Article
\bibitem[{Stieren et~al.(2021)Stieren, Gadde, and
  Stevens}]{stierenModelingDynamicWind2021}
\bibinfo{author}{A.~Stieren}, \bibinfo{author}{S.~N. Gadde},
  \bibinfo{author}{R.J.A.M. Stevens},
\newblock \bibinfo{title}{Modeling dynamic wind direction changes in large eddy
  simulations of wind farms},
\newblock \bibinfo{journal}{Renew. Energy} \bibinfo{volume}{170}
  (\bibinfo{year}{2021}) \bibinfo{pages}{1342--1352}.
  \DOIprefix\doi{10.1016/j.renene.2021.02.018}.
%Type = Article
\bibitem[{Stevens et~al.(2014)Stevens, Graham, and
  Meneveau}]{stevensConcurrentPrecursorInflow2014}
\bibinfo{author}{R.J.A.M. Stevens}, \bibinfo{author}{J.~Graham},
  \bibinfo{author}{C.~Meneveau},
\newblock \bibinfo{title}{A concurrent precursor inflow method for {{Large Eddy
  Simulations}} and applications to finite length wind farms},
\newblock \bibinfo{journal}{Renew. Energy} \bibinfo{volume}{68}
  (\bibinfo{year}{2014}) \bibinfo{pages}{46--50}.
  \DOIprefix\doi{10.1016/j.renene.2014.01.024}.
%Type = Article
\bibitem[{Stevens et~al.(2018)Stevens, {Mart{\'i}nez-Tossas}, and
  Meneveau}]{stevensComparisonWindFarm2018}
\bibinfo{author}{R.J.A.M. Stevens}, \bibinfo{author}{L.~A.
  {Mart{\'i}nez-Tossas}}, \bibinfo{author}{C.~Meneveau},
\newblock \bibinfo{title}{Comparison of wind farm large eddy simulations using
  actuator disk and actuator line models with wind tunnel experiments},
\newblock \bibinfo{journal}{Renew. Energy} \bibinfo{volume}{116}
  (\bibinfo{year}{2018}) \bibinfo{pages}{470--478}.
  \DOIprefix\doi{10.1016/j.renene.2017.08.072}.
%Type = Article
\bibitem[{Tian and Cott{\'e}(2016)}]{tianWindTurbineNoise2016}
\bibinfo{author}{Y.~Tian}, \bibinfo{author}{B.~Cott{\'e}},
\newblock \bibinfo{title}{Wind {{Turbine Noise Modeling Based}} on {{Amiet}}'s
  {{Theory}}: {{Effects}} of {{Wind Shear}} and {{Atmospheric Turbulence}}},
\newblock \bibinfo{journal}{Acta Acust. united with Acustica}
  \bibinfo{volume}{102} (\bibinfo{year}{2016}) \bibinfo{pages}{626--639}.
  \DOIprefix\doi{10.3813/AAA.918979}.
%Type = Article
\bibitem[{Amiet(1976)}]{amietNoiseDueTurbulent1976}
\bibinfo{author}{R.~Amiet},
\newblock \bibinfo{title}{Noise due to turbulent flow past a trailing edge},
\newblock \bibinfo{journal}{J. Sound Vib.}
  \bibinfo{volume}{47} (\bibinfo{year}{1976}) \bibinfo{pages}{387--393}.
  \DOIprefix\doi{10.1016/0022-460X(76)90948-2}.
%Type = Article
\bibitem[{Mascarenhas et~al.(2022)Mascarenhas, Cott{\'e}, and
  Doar{\'e}}]{mascarenhasSynthesisWindTurbine2022}
\bibinfo{author}{D.~Mascarenhas}, \bibinfo{author}{B.~Cott{\'e}},
  \bibinfo{author}{O.~Doar{\'e}},
\newblock \bibinfo{title}{Synthesis of wind turbine trailing edge noise in free
  field},
\newblock \bibinfo{journal}{JASA Express Lett.} \bibinfo{volume}{2}
  (\bibinfo{year}{2022}) \bibinfo{pages}{033601}.
  \DOIprefix\doi{10.1121/10.0009658}.
%Type = Article
\bibitem[{Lee and Shum(2019)}]{leePredictionAirfoilTrailingEdge2019}
\bibinfo{author}{S.~Lee}, \bibinfo{author}{J.~G. Shum},
\newblock \bibinfo{title}{Prediction of {{Airfoil Trailing-Edge Noise Using
  Empirical Wall-Pressure Spectrum Models}}},
\newblock \bibinfo{journal}{AIAA J.} \bibinfo{volume}{57}
  (\bibinfo{year}{2019}) \bibinfo{pages}{888--897}.
  \DOIprefix\doi{10.2514/1.J057787}.
%Type = Techreport
\bibitem[{Jonkman et~al.(2009)Jonkman, Butterfield, Musial, and
  Scott}]{jonkmanDefinition5MWReference2009}
\bibinfo{author}{J.~Jonkman}, \bibinfo{author}{S.~Butterfield},
  \bibinfo{author}{W.~Musial}, \bibinfo{author}{G.~Scott},
  \bibinfo{title}{Definition of a 5-{{MW Reference Wind Turbine}} for
  {{Offshore System Development}}}, \bibinfo{type}{Technical Report}
  \bibinfo{number}{NREL/TP-500-38060, 947422}, \bibinfo{year}{2009}.
  \DOIprefix\doi{10.2172/947422}.
%Type = Article
\bibitem[{Ostashev et~al.(2020)Ostashev, Wilson, and
  Muhlestein}]{ostashevWaveExtrawideangleParabolic2020}
\bibinfo{author}{V.~E. Ostashev}, \bibinfo{author}{D.~K. Wilson},
  \bibinfo{author}{M.~B. Muhlestein},
\newblock \bibinfo{title}{Wave and extra-wide-angle parabolic equations for
  sound propagation in a moving atmosphere},
\newblock \bibinfo{journal}{J. Acoust. Soc. Am.}
  \bibinfo{volume}{147} (\bibinfo{year}{2020}) \bibinfo{pages}{3969--3984}.
  \DOIprefix\doi{10.1121/10.0001397}.
%Type = Book
\bibitem[{Salomons(2001)}]{salomonsComputationalAtmosphericAcoustics2001}
\bibinfo{author}{E.~M. Salomons}, \bibinfo{title}{Computational Atmospheric
  Acoustics}, \bibinfo{publisher}{Kluwer Academic}, \bibinfo{address}{Dordrecht
  London}, \bibinfo{year}{2001}.
%Type = Article
\bibitem[{Cott{\'e}(2019)}]{cotteExtendedSourceModels2019}
\bibinfo{author}{B.~Cott{\'e}},
\newblock \bibinfo{title}{Extended source models for wind turbine noise
  propagation},
\newblock \bibinfo{journal}{J. Acoust. Soc. Am.}
  \bibinfo{volume}{145} (\bibinfo{year}{2019}) \bibinfo{pages}{1363--1371}.
  \DOIprefix\doi{10.1121/1.5093307}.
%Type = Article
\bibitem[{Colas et~al.(2024)Colas, Emmanuelli, Dragna, {Blanc-Benon},
  Cott{\'e}, and Stevens}]{colasImpactTwodimensionalSteep2024}
\bibinfo{author}{J.~Colas}, \bibinfo{author}{A.~Emmanuelli},
  \bibinfo{author}{D.~Dragna}, \bibinfo{author}{P.~{Blanc-Benon}},
  \bibinfo{author}{B.~Cott{\'e}}, \bibinfo{author}{R.~J. A.~M. Stevens},
\newblock \bibinfo{title}{Impact of a two-dimensional steep hill on wind
  turbine noise propagation},
\newblock \bibinfo{journal}{Wind Energy Science} \bibinfo{volume}{9}
  (\bibinfo{year}{2024}) \bibinfo{pages}{1869--1884}.
  \DOIprefix\doi{10.5194/wes-9-1869-2024}.
%Type = Article
\bibitem[{Abkar and
  {Port{\'e}-Agel}(2015)}]{abkarInfluenceAtmosphericStability2015}
\bibinfo{author}{M.~Abkar}, \bibinfo{author}{F.~{Port{\'e}-Agel}},
\newblock \bibinfo{title}{Influence of atmospheric stability on wind-turbine
  wakes: {{A}} large-eddy simulation study},
\newblock \bibinfo{journal}{Phys. Fluids} \bibinfo{volume}{27}
  (\bibinfo{year}{2015}) \bibinfo{pages}{035104}.
  \DOIprefix\doi{10.1063/1.4913695}.
%Type = Article
\bibitem[{Oerlemans and Schepers(2009)}]{oerlemansPredictionWindTurbine2009}
\bibinfo{author}{S.~Oerlemans}, \bibinfo{author}{J.~G. Schepers},
\newblock \bibinfo{title}{Prediction of {{Wind Turbine Noise}} and
  {{Validation}} against {{Experiment}}} \bibinfo{volume}{8}
  (\bibinfo{year}{2009}) \bibinfo{pages}{30}.
  \DOIprefix\doi{10.1260/147547209789141489}.
%Type = Article
\bibitem[{Mascarenhas et~al.(2023)Mascarenhas, Cott{\'e}, and
  Doar{\'e}}]{mascarenhasPropagationEffectsSynthesis2023}
\bibinfo{author}{D.~Mascarenhas}, \bibinfo{author}{B.~Cott{\'e}},
  \bibinfo{author}{O.~Doar{\'e}},
\newblock \bibinfo{title}{Propagation effects in the synthesis of wind turbine
  aerodynamic noise},
\newblock \bibinfo{journal}{Acta Acust.} \bibinfo{volume}{7}
  (\bibinfo{year}{2023}) \bibinfo{pages}{23}.
  \DOIprefix\doi{10.1051/aacus/2023018}.
%Type = Article
\bibitem[{Heimann et~al.(2018)Heimann, Englberger, and
  Schady}]{heimannSoundPropagationWake2018}
\bibinfo{author}{D.~Heimann}, \bibinfo{author}{A.~Englberger},
  \bibinfo{author}{A.~Schady},
\newblock \bibinfo{title}{Sound propagation through the wake flow of a hilltop
  wind turbine-{{A}} numerical study},
\newblock \bibinfo{journal}{Wind Energy} \bibinfo{volume}{21}
  (\bibinfo{year}{2018}) \bibinfo{pages}{650--662}.
  \DOIprefix\doi{10.1002/we.2185}.
%Type = Article
\bibitem[{Nguyen et~al.(2021)Nguyen, Hansen, Catcheside, Hansen, and
  Zajamsek}]{nguyenLongtermQuantificationCharacterisation2021}
\bibinfo{author}{P.~D. Nguyen}, \bibinfo{author}{K.~L. Hansen},
  \bibinfo{author}{P.~Catcheside}, \bibinfo{author}{C.~H. Hansen},
  \bibinfo{author}{B.~Zajamsek},
\newblock \bibinfo{title}{Long-term quantification and characterisation of wind
  farm noise amplitude modulation},
\newblock \bibinfo{journal}{Meas.} \bibinfo{volume}{182}
  (\bibinfo{year}{2021}) \bibinfo{pages}{109678}.
  \DOIprefix\doi{10.1016/j.measurement.2021.109678}.
%Type = Article
\bibitem[{{van den Berg}(2005)}]{vandenbergBeatGettingStronger2005}
\bibinfo{author}{G.~P. {van den Berg}},
\newblock \bibinfo{title}{The {{Beat}} is {{Getting Stronger}}: {{The Effect}}
  of {{Atmospheric Stability}} on {{Low Frequency Modulated Sound}} of {{Wind
  Turbines}}},
\newblock \bibinfo{journal}{Noise Notes} \bibinfo{volume}{4}
  (\bibinfo{year}{2005}) \bibinfo{pages}{15--40}.
  \DOIprefix\doi{10.1260/147547306777009247}.
%Type = Article
\bibitem[{Lee et~al.(2011)Lee, Kim, Choi, and
  Lee}]{leeAnnoyanceCausedAmplitude2011}
\bibinfo{author}{S.~Lee}, \bibinfo{author}{K.~Kim}, \bibinfo{author}{W.~Choi},
  \bibinfo{author}{S.~Lee},
\newblock \bibinfo{title}{Annoyance caused by amplitude modulation of wind
  turbine noise},
\newblock \bibinfo{journal}{Noise Control Eng. J.}
  \bibinfo{volume}{59} (\bibinfo{year}{2011}) \bibinfo{pages}{38--46}.
  \DOIprefix\doi{10.3397/1.3531797}.
%Type = Article
\bibitem[{Barlas et~al.(2018)Barlas, Wu, Zhu, {Port{\'e}-Agel}, and
  Shen}]{barlasVariabilityWindTurbine2018}
\bibinfo{author}{E.~Barlas}, \bibinfo{author}{K.~L. Wu}, \bibinfo{author}{W.~J.
  Zhu}, \bibinfo{author}{F.~{Port{\'e}-Agel}}, \bibinfo{author}{W.~Z. Shen},
\newblock \bibinfo{title}{Variability of wind turbine noise over a diurnal
  cycle},
\newblock \bibinfo{journal}{Renew. Energy} \bibinfo{volume}{126}
  (\bibinfo{year}{2018}) \bibinfo{pages}{791--800}.
  \DOIprefix\doi{10.1016/j.renene.2018.03.086}.
%Type = Article
\bibitem[{K{\"o}necke et~al.(2023)K{\"o}necke, H{\"o}rmeyer, Bohne, and
  Rolfes}]{koneckeNewBaseWind2023}
\bibinfo{author}{S.~K{\"o}necke}, \bibinfo{author}{J.~H{\"o}rmeyer},
  \bibinfo{author}{T.~Bohne}, \bibinfo{author}{R.~Rolfes},
\newblock \bibinfo{title}{A new base of wind turbine noise measurement data and
  its application for a systematic validation of sound propagation models},
\newblock \bibinfo{journal}{Wind Energy Science} \bibinfo{volume}{8}
  (\bibinfo{year}{2023}) \bibinfo{pages}{639--659}.
  \DOIprefix\doi{10.5194/wes-8-639-2023}.

\end{thebibliography}

\end{document}